\documentclass[preprint2]{aastex}

\newcommand{\ie}{{\em i.e.}}
\newcommand{\eg}{{\em e.g.}}
\newcommand{\etal}{{\em et al.}}

\newcommand{\Msun}{$M_{\odot}$}
\newcommand{\Rsun}{$R_{\odot}$}

\newcommand{\Mjup}{$M_{\rm JUP}$}

\newcommand{\sini}{$\sin i~$}
\newcommand{\msini}{$m_2 \sin i~$}
\newcommand{\vsini}{$v \sin i~$}
\newcommand{\Teff}{$T_{\rm eff}$}

\newcommand{\putFigM}[4]{}
\newcommand{\putFigT}[4]{
\begin{figure*}[!ht]
\vspace*{#1}
\includegraphics{#2}
\caption{#3}
\label{fig:#4}
\end{figure*}}

\newcommand{\putTabM}[3]{}
\newcommand{\putTabT}[3]{
\begin{table}[!ht]
\begin{center}
\caption{#2\label{table:#3}}
\vspace*{1em}
#1
\end{center}
\end{table}
}
\newcommand{\TableStellarProps}{
\begin{tabular}{lrcl}
\tableline
\tableline
Parameter & \multicolumn{3}{c}{Value} \\
\tableline
Mass [\Msun]          &  1.4  & $\pm$ & 0.09 \\
Radius [\Rsun]        &  2.14 & $\pm$ & 0.1 \\
\Teff [K]             &  6166 & $\pm$ & 145 \\
$M_v$                 &  2.78 \\
Age [Gy]              &  2.04 \\
\mbox{[Fe/H]}         &  0.18 \\
$P_{\rm rot}$ [days]  &  9   \\
\vsini [km/s]         &   8  \\
\tableline
\end{tabular}
}

\newcommand{\TableOrbParams}{
\begin{tabular}{lrcl}
\tableline
\tableline
Parameter & \multicolumn{3}{c}{Value} \\
\tableline
$P$ [day]            & 256.0 & $\pm$ &  0.7   \\
$K$ [m/s]            & 257   & $\pm$ & 14     \\
$e$                  & 0.70  & $\pm$ &  0.02  \\
$\omega$             & 195   & $\pm$ &  3     \\
$T_o$ [JD-2,450,000] & 994   & $\pm$ &  2     \\
\tableline
$f(m_1,m_2,i)$ [\Msun] & $1.64\,10^{-7}$ \\
$a_1 \sin i$ [AU]      & 0.0043 \\
\tableline
RMS(residuals) [m/s] & 20.5  \\
reduced $\chi$       &  1.6  \\
$N_{\rm Tot}$, 
$N_{\rm Rej}$        &  88,  & ~ &  1     \\
\tableline
$a_2$        [AU]    & 0.88   \\
\msini       [\Mjup] & 7.2    \\
\tableline
\end{tabular}
}

\slugcomment{Submitted to the Astrophysical Journal Letters}
\received{24 Jan 2000}
\accepted{24 Feb 2000}

\shorttitle{A Low-Mass Companion to HD\,89744}
\shortauthors{Korzennik \etal}

\begin{document}

\title{A High-Eccentricity Low-Mass Companion to HD\,89744}

\author{
 Sylvain~G.~Korzennik\altaffilmark{1},
 Timothy~M.~Brown\altaffilmark{2},
 Debra~A.~Fischer\altaffilmark{3},
 Peter~Nisenson\altaffilmark{1},
 Robert~W.~Noyes\altaffilmark{1}}

\email{skorzennik@cfa.harvard.edu}

\altaffiltext{1}{Harvard--Smithsonian Center for Astrophysics,
60 Garden St,
Cambridge, MA USA 02138}

\altaffiltext{2}{High Altitude Observatory, National
Center for Atmospheric Research, 
P.O. Box 3000, Boulder, CO USA 80307}

\altaffiltext{3}{Department of Astronomy, University of California,
Berkeley, CA USA 94720}

\begin{abstract}
  HD\,89744 is an F7 V star with mass 1.4 \Msun, effective temperature 6166~K,
age 2.0 Gy and metallicity [Fe/H]$=0.18$.  The radial velocity of the star has
been monitored with the AFOE spectrograph at the Whipple Observatory since
1996, and evidence has been found for a low mass companion. The data were
complemented by additional data from the Hamilton spectrograph at Lick
Observatory during the companion's periastron passage in fall 1999.  As a
result, we have determined the star's orbital wobble to have period $P =
256$d, orbital amplitude $K = 257$ m/s, and eccentricity $e=0.7$.  From the
stellar mass we infer that the companion has minimum mass \msini $= 7.2$
\Mjup\ in an orbit with semi-major axis $a_2 = 0.88$ AU.  The eccentricity of
the orbit, among the highest known for extra-solar planets, continues the
trend that extra-solar planets with semi-major axes greater than about 0.15 AU
tend to have much higher eccentricities than are found in our solar system.
The high metallicity of the parent star reinforces the trend that parent stars
of extra-solar planets tend to have high metallicity.
\end{abstract}


\section{INTRODUCTION}

  We report on the detection of a massive (\msini $= 7.2$ \Mjup) planet in a
highly elliptical ($e=0.7$), 256 day orbit about the star HD\,89744 (HR\,4067,
HIP\,50786), from radial velocity variations which reveal Keplerian motions of
the star.  Observations were carried out from 1996 through 1999 using the
Advanced Fiber Optic Echelle (AFOE) spectrograph \citep{BrownEtal94,
NisensonEtal98}, a bench-top spectrograph located at the Whipple 
Observatory 1.5m telescope, and also with the Hamilton spectrograph at the
Lick Observatory CAT and Shane telescopes, in November and December of 1999.

  The AFOE spectrograph is designed primarily for precise radial velocity
studies of the seismology of bright stars, and of reflex motions of stars due
to planetary companions.  Long term stability of the velocity reference is
provided by use of an iodine ($I_2$) cell \citep{ButlerEtal96}.  The AFOE
determines radial velocity variations induced by planetary companions with a
precision and long-term accuracy of approximately 10 m/s.  On the order of 100
relatively bright stars ($m_v \leq 7$) have been monitored for this purpose
since 1995.

  Since 1995 when the planetary candidate oribiting the star 51 Pegasus was
detected \citep{MayorQueloz95}, some 29 additional candidates have been
detected by several groups, all from Doppler shifts measured using precise
radial velocity techniques \citep{MarcyButler98, MayorEtal98, NoyesEtal97,
CochranEtal97}.

  HD\,89744 (F7 V) was added to the AFOE observing list in early 1996, based
on its relatively low chromospheric emission as measured with the Mt.\ Wilson
``HK'' chromospheric activity monitoring program \citep{BaliunasEtal95}.  AFOE
observations have been obtained regularly since then, and indicated the
presence of a planet with a highly eccentric orbit.  However data near the
companion's periastron, critical to an accurate determination of the orbital
parameters, were not obtained until late 1999.  Between October and December
1999, while the companion was near periastron, observations were made at Lick
Observatory as well as with the AFOE, to ensure good phase coverage.  The data
points taken with the Lick CAT and Shane telescopes agree extremely well with
the AFOE data, and thus provide a confirmation of the detection along with a
precise determination of the ellipticity of the planet's orbit.

\section{PROPERTIES OF THE HOST STAR, HD\,89744}

  HD\,89744 is an F7V star at a Hipparcos-determined distance of 39.0 parsec.
It is listed as a constant star in the Hipparcos catalog . The star has
absolute magnitude $M_v = 2.78$ and color $B-V=0.531$
\citep{Perryman97}. Comparing its position in the color-magnitude diagram with
predictions of stellar evolution calculations, \citet{Prieto+Lambert00}
determine it to have mass $M=(1.34 \pm 0.09) M_{\odot}$, radius $R=(2.14 \pm
0.1) R_{\odot}$, and effective temperature \Teff $ = (6166 \pm 145) K$.
Independently, \citet{Ng+Bertelli98} determine its mass to be $M=(1.47 \pm
0.01) M_{\odot}$.  For the purposes of this paper we adopt the average of the
two masses listed and an uncertainty given by their spread: $M=(1.4 \pm 0.09)
M_{\odot}$.

  The metallicity of HD\,89744 has been determined to be [Fe/H]$= 0.18$ by
\citet{EdvardssonEtal93}.  Its age is determined by \citet{Ng+Bertelli98} to
be $2.04 \pm 0.10$ Gy.

  From rotational modulation of the \ion{Ca}{2} flux, \citet{BaliunasEtal96}
determined that the rotation period of the star is $P_{\rm rot} = 9$ days.
This, combined with the above-mentioned radius of the star, implies an
equatorial velocity $v$ = 12 km/sec.  Rotational broadening of the spectrum
implies \vsini $ = 8$ km/s \citep{Bernacca+Perinotto70, Uesugi+Fukuda70}.
This further implies that the star's rotational equator is inclined by about
$40^{\circ}$\ to the plane of the sky.

  Speckle observations \citep{McAlisterEtal89} have not revealed any
indication of a stellar companion to HD\,89744.

  Table~\ref{table:stellarProps} summarizes the relevant parameters of the
star HD\,89744.

\putTabT{\TableStellarProps}{Parameters for HD\,89744.}{stellarProps}

\section{OBSERVATIONS}
\subsection{Instrumentation and Data Reduction}

  The AFOE is a bench-mounted, fiber-fed, high-resolution cross-dispersed
echelle spectrograph designed for high radial velocity precision and stability
both on short time scales (better than 1 m/s over hours, for asteroseismology)
and long term (approximately 10 m/s, for radial velocity exo-planet searches,
through use of an iodine absorption cell). The AFOE is located at the 1.5m
telescope of the Whipple Observatory on Mt.\ Hopkins, Arizona.  A more
complete description of the AFOE is available in \cite{BrownEtal94} and
\cite{NisensonEtal98}.  The AFOE exo-planet survey program monitors the radial
velocity of about 100 stars brighter than $m_v = 7.5$, with an accuracy of 10
-- 15 m/s for integrations with a signal-to-noise ratio of 100 to 150.  Most
observations consist of three consecutive exposures, primarily to limit cosmic
ray contamination by keeping the exposure times short.

  Our data reduction methodology is conceptually similar to that described by
\citet{ButlerEtal96}, but differs in details. Echelle images are dark
subtracted and one dimensional spectra extracted and corrected for scattered
light, then flat-fielded using the spectrum of a tungsten lamp. For each of
the six spectral orders that contain strong iodine lines a model is adjusted
to match the observed star-plus-iodine spectrum in the least-squares
sense. This model is computed using a Doppler-shifted high SNR spectrum of the
star alone, plus a very high resolution high SNR spectrum of the iodine
cell. The model incorporates the sought-for Doppler-shift of the star as well
as mechanical drifts within the spectrograph, the instrumental wavelength
solution, the instrumental resolution profile, and a residual scattered light
correction.  The resulting radial velocities, after correction for the motion
of the telescope relative to the solar system barycenter, are averaged for all
six orders and the three consecutive observations.  The scatter around the
mean provides an estimate of the uncertainty. The root-mean-square (RMS)
velocity of several radial velocity standard stars is commensurate with this
estimate of uncertainty.  The Doppler analysis for the Lick observations is
described in detail in \cite{ButlerEtal96}.

\subsection{Observations and Orbital Fit}

AFOE observations of HD\,89744 were obtained on 74 separate nights between
December 1996 and December 1999. Additional observations on 14 nights during
November and December 1999 were also taken at the Lick Observatory CAT and
Shane telescopes, to ensure complete phase coverage near the November 1999
periastron passage. The data with their uncertainties are plotted in
Figure~\ref{fig:rawData}.

\putFigT{4in}{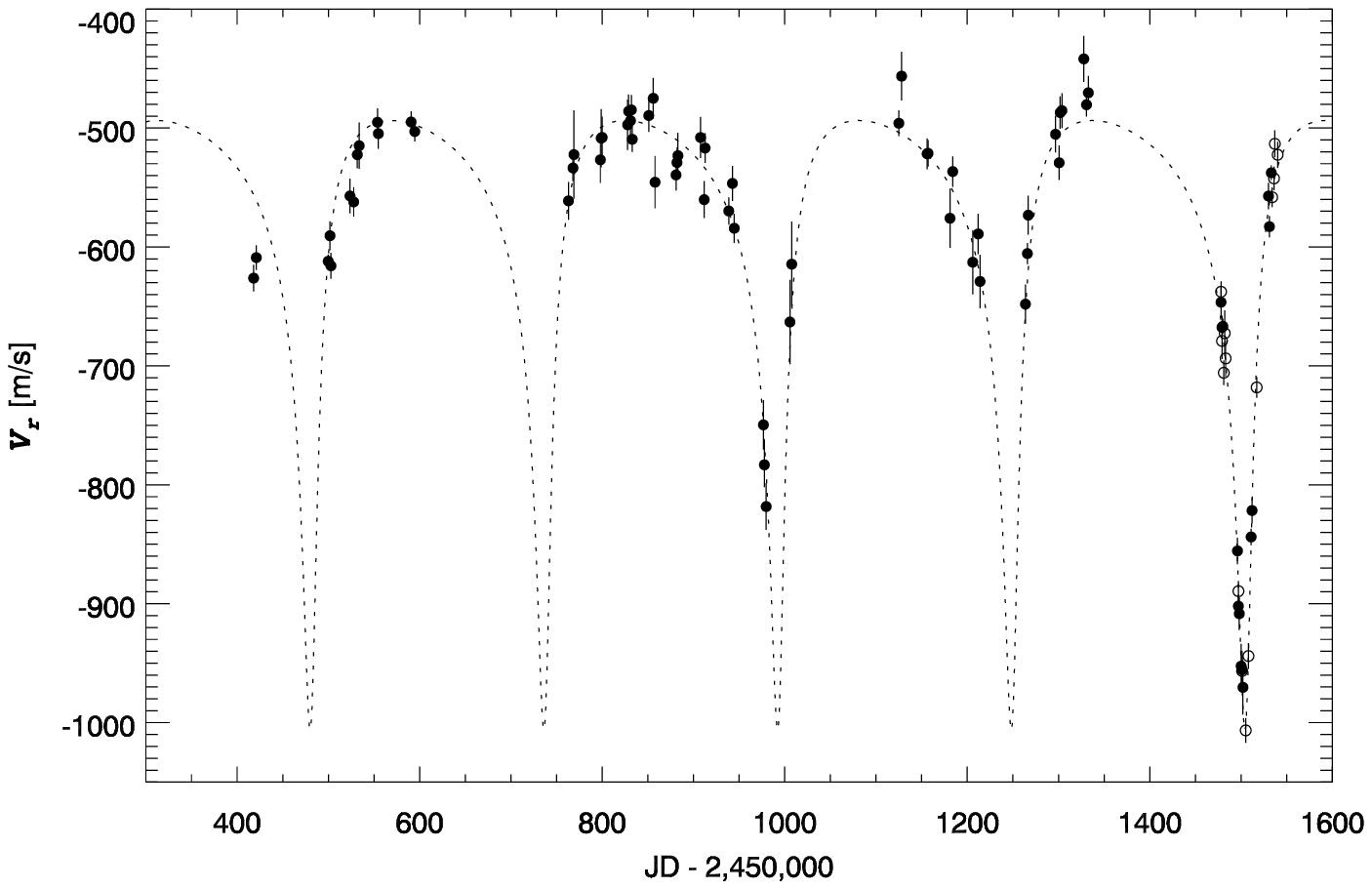}
{Radial velocity observations of HD\,89744 with their respective
uncertainties. Filled circles denote AFOE observations, and open circles Lick
observations after applying an offset determined by the combined orbital
fit. The Keplerian orbital fit is indicated by the dotted line.}
{rawData}

 The RMS of the zero-averaged velocities for the AFOE data alone is 129.1 m/s,
while the averaged uncertainty on these observations is 15.2 m/s; the
corresponding reduced $\chi$ is 10; where the reduced $\chi$ is given by
\begin{equation}
  \chi = \left[\frac{1}{N_o-N_m} 
               \sum_{i=1}^{N_o}
              (\frac{O_i-M_i}{\sigma_i})^2\right]^{\frac{1}{2}}
\end{equation}
where $O_i$ are the observations, $\sigma_i$ the observational uncertainties,
$M_i$ the corresponding values of the model, $N_o$ the number of observations
and $N_m$ the number of adjustable parameters in the model.
  After fitting a Keplerian orbit in the
least-squares sense, the RMS of the residuals is 26.2 m/s, corresponding to a
reduced $\chi$ of 1.7.  If we include the Lick observations, by adding one
additional free parameter to account for the arbitrary offset between the two
data sets, the RMS of the residuals is 20.5 m/s, or a reduced $\chi$ of 1.6.
(One velocity point was rejected based on a 3-pass 3-$\sigma$ rejection
algorithm.) The resulting orbital parameters are given in Table 2. These are
almost identical to the parameters obtained when using only the AFOE
observations, except that the combined data set leads to smaller
uncertainties, primarily on the amplitude but also on the eccentricity. The
orbital fit phase plot and the residuals to the fit are shown in
Figures~\ref{fig:phasePlot} and \ref{fig:residuals}. The periodic variation of
the radial velocity, and the highly eccentric character of the orbit, are
evident.  For a stellar mass of 1.4 \Msun, the orbital elements imply a
minimum mass for the companion of \msini $ = 7.2$ \Mjup, and a semi-major
axis of 0.88 AU.

\putTabT{\TableOrbParams}{Orbital parameters resulting from a Keplerian fit to
the combined data set.  The last two rows of the table use the adopted value
$m_1 = 1.4$ \Msun}{orbParams}

\putFigT{4in}{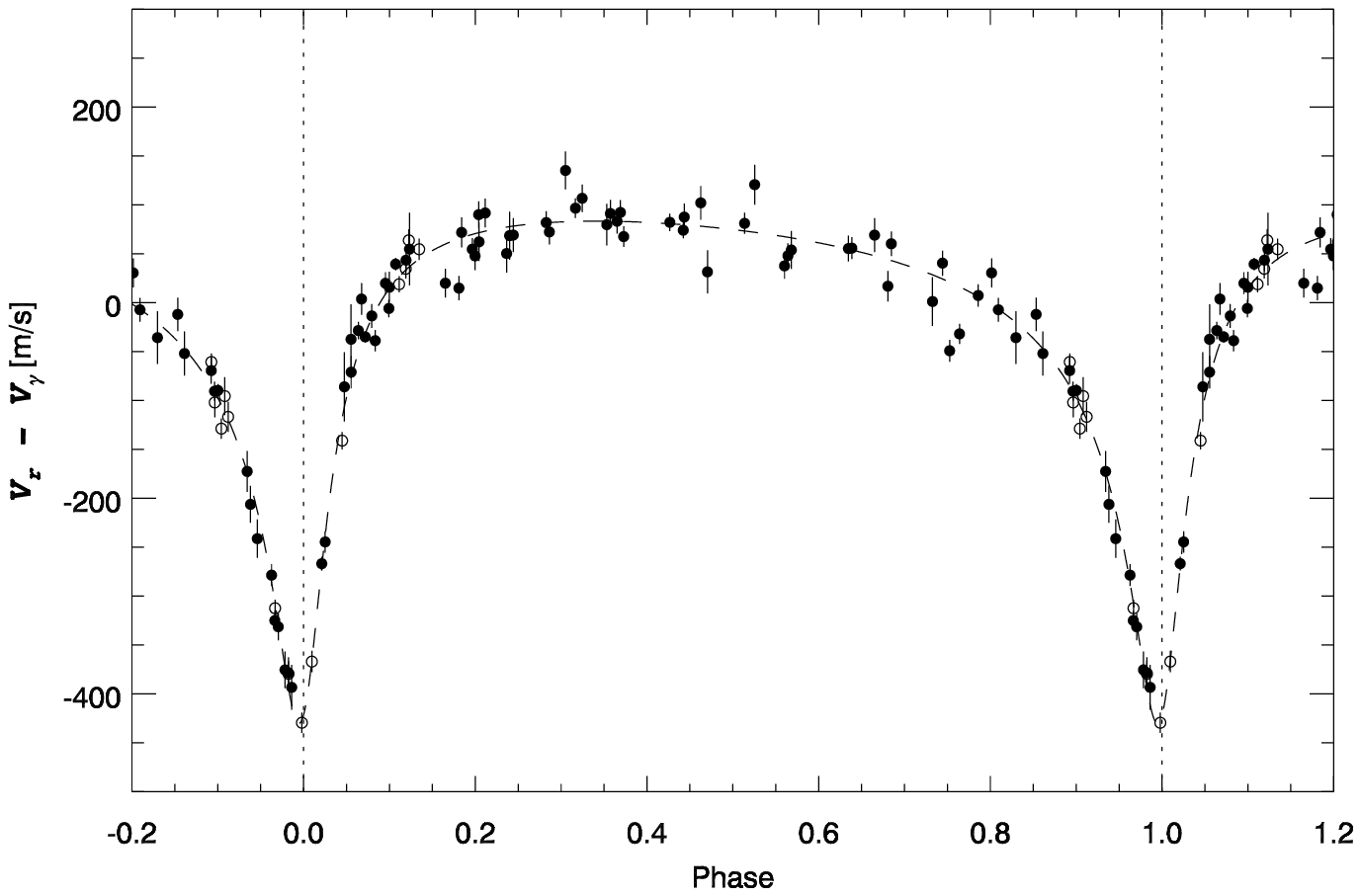}
{Phase plot of the Keplerian orbital fit. Filled circles denote AFOE
observations, open circles Lick observations.}
{phasePlot}

\putFigT{4in}{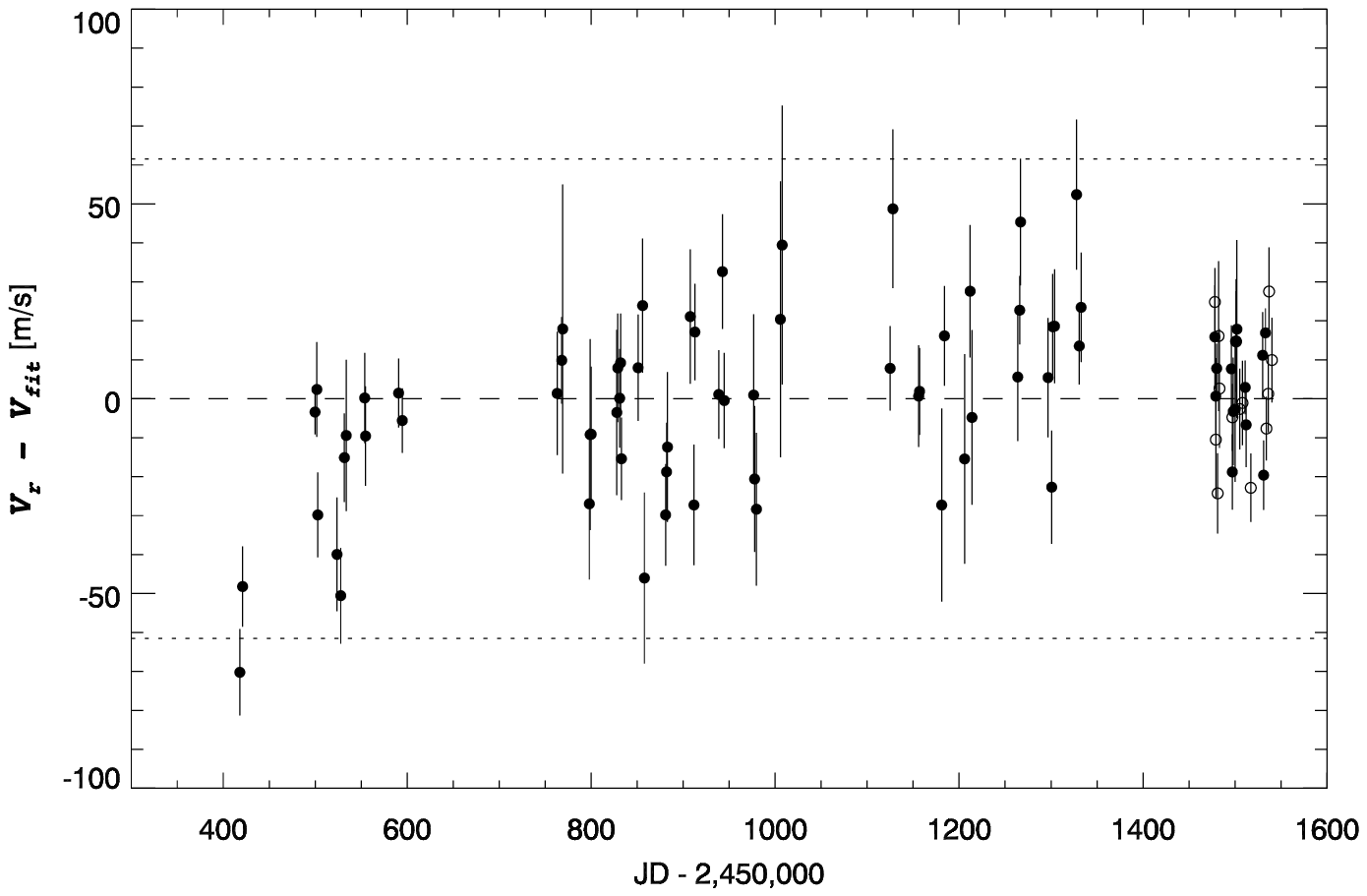}
{Residuals to the Keplerian orbital fit. Filled circles denote AFOE
observations, open circles Lick observations. The dotted lines illustrate the
3-sigma rejection bounds.}
{residuals}

\section{DISCUSSION}

The radial velocity data shown in Figure~\ref{fig:phasePlot} are unambiguous
in revealing a periodic radial velocity variation of HD\,89744, which can be
fit well by a Keplerian orbit.  Since observations obtained with two different
radial velocity instruments at two different telescopes yield exactly the same
radial velocity variations within their respective instrumental errors, the
evidence is compelling that the measured velocities are real variations on the
star. There is no known way a stellar signal in a late F-type main sequence
star could mimic a Keplerian orbital signature with such a long period and
large amplitude, and with such a large eccentricity.  Hence we are driven to
the interpretation that the star is orbited by a low-mass companion,
HD\,89744~b, (\msini $= 7.2 M_{\odot}$), in an orbit with semi-major axis 0.88
AU and eccentricity 0.70.

  The residuals to the orbital fit are larger than would be expected from
internal errors in the data; this is true both for the AFOE data and the
Hamilton echelle data.  However, \citet{SaarEtal98} conclude that for a
typical F stars with this rotation period, the velocity jitter induced by
stellar magnetic activity and inhomogeneous convection is approximately 10
m/s. Adding this jitter in quadrature to the uncertainties lower the reduced
$\chi$ to 1.2.  The residuals display a long term trend ($\approx 15$
m/s/year) that is marginally significant. While it might be caused by a
residual instrumental drift in the AFOE data, we can not rule out that it
might be due to a distant companion; further observations over a longer
baseline are required.

  The orbital eccentricity of the companion to HD\,89744 is among the highest
planetary eccentricities known. Only two other planets, 16\,Cyg\,B~b \citep[$e
= 0.68$, $a_2 = 1.70$ AU;][]{CochranEtal97} and HD\,222582~b \citep[$e =
0.71$, $a_2 = 1.35$ AU;][]{VogtEtal99} have comparable eccentricities.
HD\,89744~b, with $a_2 = 0.88$ AU, has the smallest semi-major axis of the
three.  At periastron it dips to within 0.26 AU, still well outside a
periastron distance which could lead to tidal circularization within the
stellar lifetime.

  The discovery of the first highly eccentric planet, 16\,Cyg\,B~b, led to the
suggestion \citep{MazehEtal97, HolmanEtal97} that its eccentricity may
have been ``pumped up'' by the influence of a nearby companion star,
16\,Cyg\,A.  However, neither HD\,89744 nor HD\,222582 is orbited by a stellar
companion.  Thus a different explanation is required for their high
eccentricities, an explanation that might also apply to 16\,Cyg\,B~b.

  The high orbital eccentricity of the HD\,89744 system continues the trend
that planetary-mass companions whose semi-major axes are greater than about
0.15 AU tend to have a broad range of eccentricities, with no apparent trends
of eccentricity with mass or semi-major axis. This circumstance must be
explained by any successful planetary formation and migration scenario. As
noted by others \citep[\eg,][]{VogtEtal99, MarcyEtal99}, this causes
difficulties with a number of proposed mechanisms for planetary formation and
migration.

  The values of $v$sin$i$, $P_{\rm rot}$, and $R$ for HD\,89744 given in
Table~\ref{table:stellarProps} imply an inclination of the stellar rotational
equator of $i= 42^{\circ}$ (\ie, \sini = 0.66). Moreover, if we assume that
the orbit is coplanar with the star's equatorial plane, we infer that $m_2$ =
10.8 \Mjup.  The astrometric orbital amplitude would be 0.17 mas, too small
for Hipparcos detection but within the range of next-generation astrometric
missions.

  The mass of 10 \Mjup\ suggested by the stellar rotational data is near the
upper limit of masses associated with extra-solar giant planets
\citep[\eg,][]{MarcyEtal00}.  If such large masses hold up to further
investigations, then theoretical understanding of the origin and evolution of
extra-solar giant planets must be able to accommodate a mass range spanning at
least values between 0.5 \Mjup\ and 10 \Mjup.

  The metallicity of HD\,89744, [Fe/H]= 0.18, is substantially higher than the
mean for nearby sun-like stars \citep{FavataEtal97, Gonzalez98}.  HD\,89744
was placed on the AFOE observing list without reference to its metallicity;
therefore the association of its high metallicity with the presence of a
planet is not a selection effect.  This association continues the trend,
already noted \citep[\eg,][ and references therein]{Gonzalez+Laws99} and
references therein that the metallicity of stars with planets tends to be
higher than that of stars without planets.

\acknowledgements
We are grateful to the Mt.\ Hopkins observing and support staff, especially Ed
Bennet, Perry Berlind, Mike Calkins, Ted Groener, Robert Hutchins, Jim Peters
and Wayne Peters; we also acknowledge observing with the AFOE by Scott Horner
and Ted Kennelly. We are very grateful to Adam Contos for his dedicated
efforts in observing and reducing a significant fraction of data taken with
the AFOE. We also thank Ari Behar for his help in reducing some of the AFOE
data.  DAF acknowledges the dedication and pioneering efforts of G.\ Marcy and
P.\ Butler who started the Lick planet search project. She would also like to
thank Sabine Frink, David Nidever and Amy Reines for obtaining some of the
Lick observations.  The AFOE group also acknowledges support from NASA grant
NAG5--75005 and from the Smithsonian Institution Scholarly Studies program.
DAF acknowledges support from NASA grant NAG5--8861.  In preparation of this
paper, we made use of the Simbad database operated at CDS, Strasbourg, France
and the NASA Astrophysics Data System.

%
%%%   repeat table and figs for manuscript mode   %%%
%%% ** must be identical to previous reference ** %%%
%
\putTabM{\TableStellarProps}{Parameters for HD\,89744.}{stellarProps}
\putTabM{\TableOrbParams}{Orbital parameters resulting from a Keplerian fit to
the combined data set.  The last two rows of the table use the adopted value
$m_1 = 1.4$ \Msun}{orbParams}
\putFigM{4in}{plot1.ps}
{Radial velocity observations of HD\,89744 with their respective
uncertainties. Filled circles denote AFOE observations, and open circles Lick
observations after applying an offset determined by the combined orbital
fit. The Keplerian orbital fit is indicated by the dotted line.}
{rawData}
\putFigM{4in}{plot2.ps}
{Phase plot of the Keplerian orbital fit. Filled circles denote AFOE
observations, open circles Lick observations.}
{phasePlot}
\putFigM{4in}{plot3.ps}
{Residuals to the Keplerian orbital fit. Filled circles denote AFOE
observations, open circles Lick observations. The dotted lines illustrate the
3-sigma rejection bounds.}
{residuals}
\end{document}